\documentclass[twocolumn,aps,graphicx]{revtex4}

\usepackage{graphicx}
\usepackage{dcolumn}
\usepackage{bm}

\begin{document}
\title{Ultracold Collisions of Fermionic OD Radicals}
\author{A. V. Avdeenkov and John L. Bohn}
\affiliation{JILA and Department of Physics,
University of Colorado, Boulder, CO 80309-0440}
\date{\today}

\begin{abstract}
We discuss consequences of Fermi exchange symmetry on collisions
of polar molecules at low temperatures (below 1 K), considering
the OD radical as a prototype.  At low fields and low temperatures,
Fermi statistics can stabilize a gas of OD molecules against
state-changing collisions.  We find, however, that this stability 
does not extend to temperatures high enough to assist with 
evaporative cooling.  In addition, we establish that a novel
``field-linked'' resonance state of OD dimers exists, in analogy with
the similar states predicted for bosonic OH.
\end{abstract}

\maketitle

\section{Introduction}

Cooling and trapping molecules in their ro-vibrational
ground states has proven to
be a daunting experimental task, yet now it has been
achieved \cite{Weinstein,Bethlem,Meerakker}.  Currently, the 
samples produced by Stark slowing are limited to temperatures around
1 mK, which is cold enough to trap, but not yet cold enough
for interesting applications to novel dilute quantum gases of
fermions~\cite{You,Goral1,Baranov1,Baranov2,Baranov3}
or bosons~\cite{Yi1,Santos,Yi2,Goral2,Giovanazzi,Odell}.
  To produce colder, denser samples for these applications, 
an attractive approach may be to use sympathetic
cooling with an easily cooled species (such as Rb) \cite{Soldan},
or else evaporative cooling.  Knowledge of collision cross sections
is therefore essential in understanding prospects for the
success of either approach.

We have previously considered electrostatic trapping of polar 
$\Pi$-state molecules 
from the point of view of stability with respect to collisions
\cite{Avdeenkov1}. 
The main bottleneck here is that electro{\it static} trapping requires
the molecules to be in a weak-field-seeking state, in which case the molecules
of necessity have an even lower-energy strong-field seeking state.  Collisions
involving the strong and anisotropic dipole-dipole interaction between
molecules appear more than adequate to drive the molecules into these
unfavorable states, leading to unacceptably high trap loss and
heating.
For this reason, it may be necessary to seek alternative methods
that can confine the strong-field-seeking, absolute lowest-energy ground
state of the molecule using time-varying electric 
fields~\cite{Cornell,Spreeuw,Junglen,DeMille}.
A recent proposal for an electro{\it dynamic} trap
is based on the microwave analogue of the familiar far-off-resonant
optical dipole trap -- only the microwave version doesn't have to be
far off-resonance, making the trap very deep \cite{DeMille}.

Polar fermions may have an important advantage  for 
electrostatic trapping, namely, low inelastic rates at cold temperatures.
  Kajita has discussed state-changing collisions
of dipolar fermionic molecules \cite{Kajita}, based
 on the well-known Wigner threshold laws for dipolar interactions.
Namely, elastic scattering cross sections are essentially
independent of collision energy $E$ at low energies, but state-changing
cross sections scale as $E^{1/2}$.  Therefore, at ``sufficiently low''
temperatures, elastic scattering always wins, and evaporative cooling
should be possible.  Using the 
 Born approximation, Kajita concludes 
that this is the case for the molecules OCS and CH$_3$Cl,
at reasonable experimental temperatures \cite{Kajita}.  This analysis 
may yet prove too optimistic, since the results include regions where 
the Born approximation may not be strictly applicable \cite{Avdeenkov2}. 
Still, the idea is a sound one that deserves further exploration.
 
A complete theoretical description of molecule-molecule scattering
is complicated by the complexity of the short-range interaction
between molecules.  Indeed, for open shell molecules the
potential energy surface is difficult to compute by {\it ab intio}
methods, and remains inadequately known.  It is therefore worthwhile
to seek situations in which the influence of short-range physics is minimal.
It appears that for weak- field seeking states
the influence of the short-range potential is weak, owing to
avoided crossings in the long-range interaction~\cite{Avdeenkov1}.
For collisions of identical fermionic molecules, the influence of
short-range physics may be even smaller, since only partial waves
with $l \ge 1$ are present, and there is centrifugal repulsion
in all scattering channels.

A main aim of the present paper is thus to explore the suppression of inelastic
collisions in fermionic $\Pi$-state molecules, using the OD radical as an
example.  This is an illustrative choice of molecule, since we have studied
its bosonic counterpart, OH, extensively in the past 
\cite{Avdeenkov1,Avdeenkov3,Avdeenkov4}.  It is also a species
at the center of current experimental interest \cite{Meerakker,Bochinski}.
To this end we employ
full close-coupling calculations to a model of the OD-OD interaction that
includes only the dipolar part.  We find, as we must, that the fermionic
threshold laws ultimately favor elastic over inelastic scattering 
at low temperatures.  For OD, however, we find that the energy 
scales for this to happen remain quite low, on the order of microKelvins 
or below, so that the usefulness of this result
to evaporative cooling remain questionable.  

On the bright side, the suppression of inelastic collisions does mean that
a gas that is already cold will be stable under collisions, even in an 
electrostatic trap.  This is a similar conclusion to one we have drawn 
in the past for magnetostatic trapping of spin-polarized paramagnetic 
(nonpolar) species \cite{Avdeenkov5,Volpi}.  This is useful
for cold collisions studies, since it is believed that collisions 
of weak-electric-field seekers are dominated by, and can be 
understood in terms of, purely long-range dipolar forces \cite{Avdeenkov3}.
In particular, such collisions are predicted for bosons
to have long-range resonant states, termed field-linked resonances, that may
be useful in understanding cold collisions.  A second goal of this paper
is to verify that the fermionic OD molecules also possess these resonances.

\section{Threshold laws in the Born approximation}

Threshold laws for various power-law long-range potentials have
been written about extensively in the past 
\cite{Wigner,Landau,Spruch,OMalley,Hinckelman,Shakeshaft,Holzwarth,Peach,Fabrikant,Rosenberg,Cavagnero,Gao,Sadeghpour,Deb}.
In this
section we summarize the main results relevant to the energy
dependence of cross sections, using the first Born approximation to
make the math transparent.  Similar arguments are presented
in Refs.~\cite{Yi1,Kajita}.

A first point to be considered is why the Born approximation should
be of any use at all, since it is ordinarily associated with collisions
of ``fast'' particles.  Strictly speaking, however, the Born approximation
is valid when the potential responsible for scattering is suitably
``weak,'' meaning that the true scattering wave function is 
well-approximated by the unperturbed wave function.  For dipolar
scattering, the argument is as follows.  Consider elastic scattering
in a single-channel whose long-range potential varies as $1/R^s$.
Then, in partial wave $l$, the elastic scattering phase shift
will vary with wave number $k$ as \cite{Sadeghpour}
\begin{equation}
\label{delta}
\delta_l \sim \alpha k^{2l+1} + \beta k^{(s-1)},
\end{equation}
where $\alpha$ and $\beta$ are constants depending on details
of the potential.

Thus for a dipolar potential with $s=3$, the second term in
Eqn.~(\ref{delta}) is the dominant contribution to the phase
shift for any partial wave $l \ge 1$, yielding $\delta_l \sim k$.  
(Moreover, the $l=0$ contribution to a realistic dipole-dipole
interaction rigorously vanishes by symmetry.)  It can be shown
(for example, using the JWKB approximation~\cite{Sadeghpour}) that the 
second contribution in Eqn.~(\ref{delta}) arises from purely
long-range physics, i.e., for intermolecular separations
outside the centrifugal barrier imposed by the partial wave.
As the collision energy approaches threshold, this distance
gets ever larger, and the influence of the $1/R^3$ perturbing
potential gets ever weaker.  Thus, near threshold, the wave function
is well-approximated by unperturbed spherical Bessel functions
in each partial wave, and the Born approximation can be used.

We adopt this view in the multichannel case.  Because the
dipole-dipole interaction is anisotropic, different partial
waves are coupled together.
Nevertheless, the diagonal pieces of the Hamiltonian matrix
have the general form
\begin{equation}
\label{potential}
{\hbar ^2 l(l+1) \over 2 \mu^2 R^2} + { C_{3}^{\rm eff} \over R^3},
\end{equation}
where $R$ is the distance between molecules, $\mu$ is their reduced mass, 
and the effective $C_3$ coefficient depends, in general,
on the channel as well as on the degree of 
electric field polarization (see Ref.\cite{Avdeenkov1}).
When $C_3^{\rm eff}$ is negative, the potential (\ref{potential})
presents a finite barrier of height 
\begin{equation}
E_b = {4 \over 27} \left[ {\hbar^2 l(l+1) \over 2 \mu} \right]^3
{1 \over (C_3^{\rm eff})^2 }.
\end{equation}
For energies $E$ considerably less than $E_b$, scattering only 
occurs from outside the barrier (barring resonances~\cite{Yi1}), thus
setting an energy scale for the utility of the Born approximation.
To make an estimate of this energy scale, consider the strong-field
limit, where polarized molecules have $C_3 \sim d^2$, the square
of the dipole moment.  For OD, this sets the relevant p-wave
centrifugal barrier height at $\sim 10$ nK.  At higher energies,
the incoming wave spills over the barrier, samples smaller-$R$
interactions, and is no longer well-described as a plane wave.

Assuming that the Born approximation holds, we proceed as follows.
The partial scattering cross section for a collision process entering on
channel $i$ and exiting on channel $f$ is given in terms of 
the transition matrix $T$ by
\begin{equation}
\label{cross}
\sigma_{if} = {\pi \over k_i^2} |\langle i | T | f \rangle|^2,
\end{equation}
where the channel indices $i$ and $f$ include partial wave
contributions $l_i$ and $l_f$, which need not be the same.
In the first Born approximation, the $T$ matrix elements are given
by the matrix elements of the potential (Chap. 7 of Ref.~\cite{Child},
where we have re-inserted the dimensionful factors)
\begin{eqnarray}
\label{Born}
\langle i | T | f \rangle = && 2 \left({2 \mu \over \hbar^2}\right)
\left( k_i k_f \right)^{1/2}  \\
& & \times \int_0^{\infty} R^2 dR j_{l_i}(k_iR) 
{C_3(l_il_f) \over R^3} j_{l_f}(k_fR) . \nonumber
\end{eqnarray}
Here $C_3(l_il_f)$ represents the appropriate off-diagonal 
coupling matrix element, which, again, depends on field.

For {\it elastic} scattering, where the initial and final wave numbers
are equal, $k_f=k_i$, we can rewrite Eqn.~(\ref{Born}) in terms of
the dimensionless variable $x=k_iR$,
\begin{eqnarray}
\label{Bornelastic}
\langle i | T | f \rangle = && {4 \mu C_3(l_il_f) \over \hbar^2} k_i \\
&& \times \int_0^{\infty} dx {j_{l_i}(x) j_{l_f}(x) \over x}. \nonumber
\end{eqnarray}
The integral in Eqn.~(\ref{Bornelastic}) converges whenever
$l_i + l_f > 0$, and is moreover independent of $k_i$.  Therefore,
for any elastic scattering process by dipolar forces that changes
$l$ by at most 2 units, $T \sim k_i$ at low energies, and by 
Eqn.~(\ref{cross}), the associated cross section is independent
of collision energy.  In particular, the elastic scattering cross
section of identical fermions does not vanish, if they interact
via dipolar forces.

For completeness, we give the value of the integral.  This is found by
substituting ordinary Bessel functions for the spherical Bessel
functions, $j_n(x) = \sqrt{ \pi /2x} J_{n+1/2}(x)$, and using
standard formulas for integrals \cite{Gradshteyn}:
\begin{eqnarray}
\int_0^{\infty} && dx {j_{l_i}(x) j_{l_f}(x) \over x}  \\
=&& { \pi \Gamma ({l_i+l_f \over 2}) \over 
8 \Gamma ({-l_i+l_f+3 \over 2}) \Gamma ({l_i+l_f+4 \over 2})
\Gamma ({l_i-l_f+3 \over 2}) }. \nonumber
\end{eqnarray}

For an {\it exothermic} process, with $k_f > k_i$, a similar argument
yields for the transition amplitude
\begin{eqnarray}
\label{Borninelastic}
\langle i | T | f \rangle  &&= \pi {2 \mu C_3(l_il_f) \over \hbar^2} \\
&& \times \int_0^{\infty} dR
J_{l_i+1/2}(k_iR) J_{l_f+1/2}(k_fR) R^{-2}. \nonumber
\end{eqnarray}
This integral, too, can be done as long as $l_i+l_f>0$
\cite{Gradshteyn}:
\begin{eqnarray}
\label{inelastic}
&& \int_0^{\infty} dR J_{l_i+1/2}(k_iR) J_{l_f+1/2}(k_fR) R^{-2} \\
\nonumber
&& = {k_i^{l_i+1/2} \Gamma( {l_i+l_f  \over 2}) \over
4 k_f^{l_i-l_f+1} \Gamma ({ -l_i+l_f+3 \over 2} ) \Gamma( l_i+3/2)} \\
\nonumber
&& \times F \left( {l_i+l_f  \over 2}, {l_i-l_f-1 \over 2},
l_i+3/2; \left( {k_i \over k_f} \right)^2 \right), \nonumber
\end{eqnarray}
where $F$ stands for a hypergeometric function.

Near threshold in an exothermic process, we have $k_f \gg k_i$.
In this case the leading order term of the hypergoemetric function
$F$ is a constant, and the only remaining dependence of
(\ref{inelastic}) on $k_i$ is in its prefactor.  Thus
$T \sim k_i^{l_i+1}$, and $\sigma \sim k_i^{2l_i-1} \sim
E^{l_i-1/2}$.
When the incident partial wave is $l_i = 0$, as would be the case
for identical bosons, the inelastic scattering cross section
diverges at threshold.  For any higher partial wave, say the 
$l_i=1$ partial wave that dominates scattering of identical
fermions, the inelastic cross section instead vanishes in the
threshold limit.

\section{Collision cross sections for OD}

The OD radical differs from OH in two significant ways, for our present 
purposes: first, its lambda-doubling
constant is somewhat smaller \cite{Abrams} .  Second, its hyperfine structure
depends on the nuclear spin of deuterium being 1 instead of 1/2 for hydrogen,
meaning that total spin states $f=1/2, 3/2$, and 5/2 are possible in 
the $^2\Pi_{3/2}$ electronic ground state of OD (as in our OH work, 
we consider exclusively in the electronic
ground state, and neglect excited vibrational and rotational levels). 
Figure 1 presents the Stark effect for OD, which can be compared to
the similar figure for OH, Fig. (1) of Ref.\cite{Avdeenkov1}.  
Note that, due to the smaller lambda-doublet in OD, this radical 
enters the linear Stark regime at applied electric fields of 
$\sim 200$ V/cm, as opposed to $\sim 1000$ V/cm in OH.  

\begin{figure}
\centerline{\includegraphics[width=0.9\linewidth,height=1.08\linewidth,angle=-90]{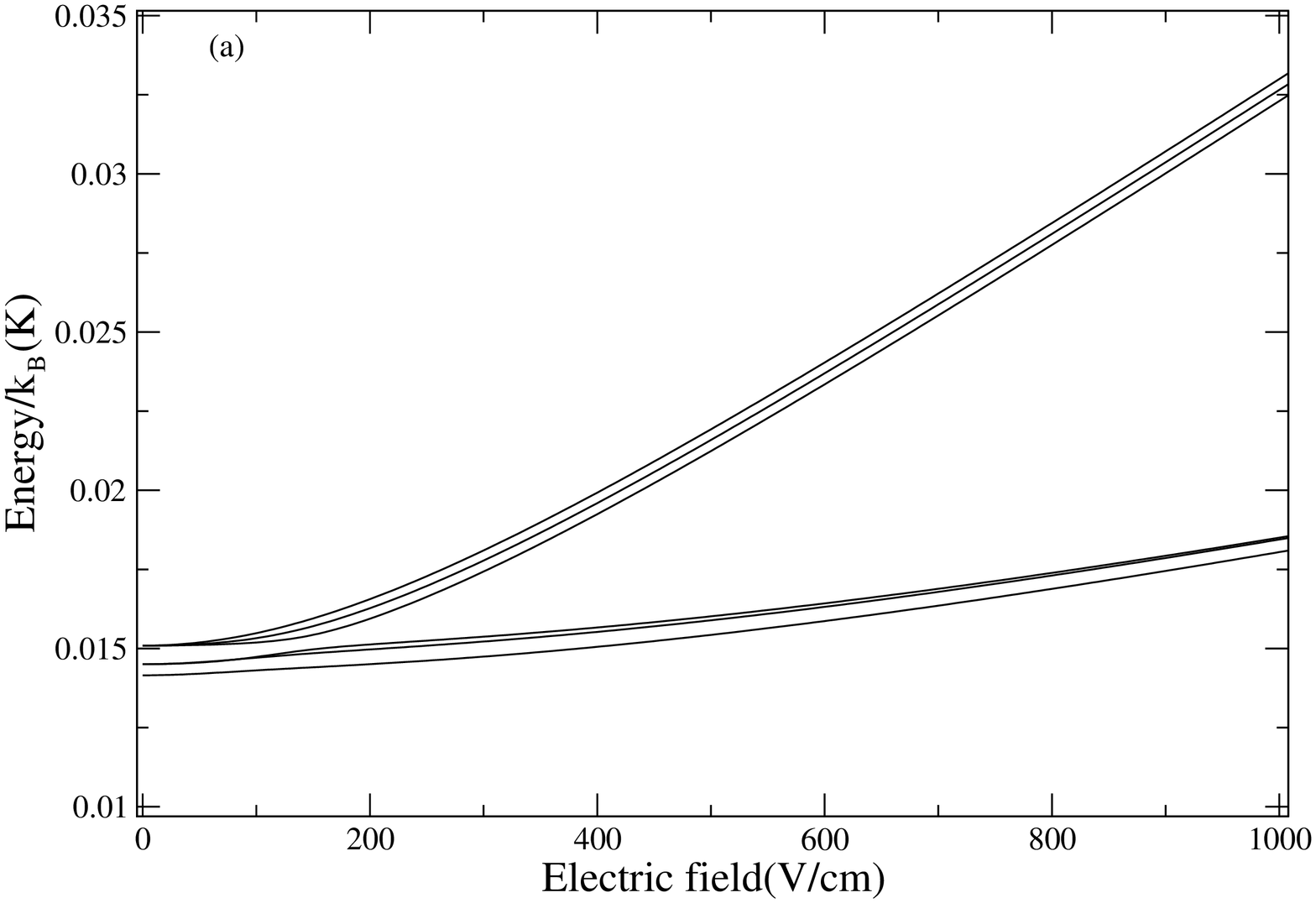}}
\centerline{\includegraphics[width=0.9\linewidth,height=1.08\linewidth,angle=-90]{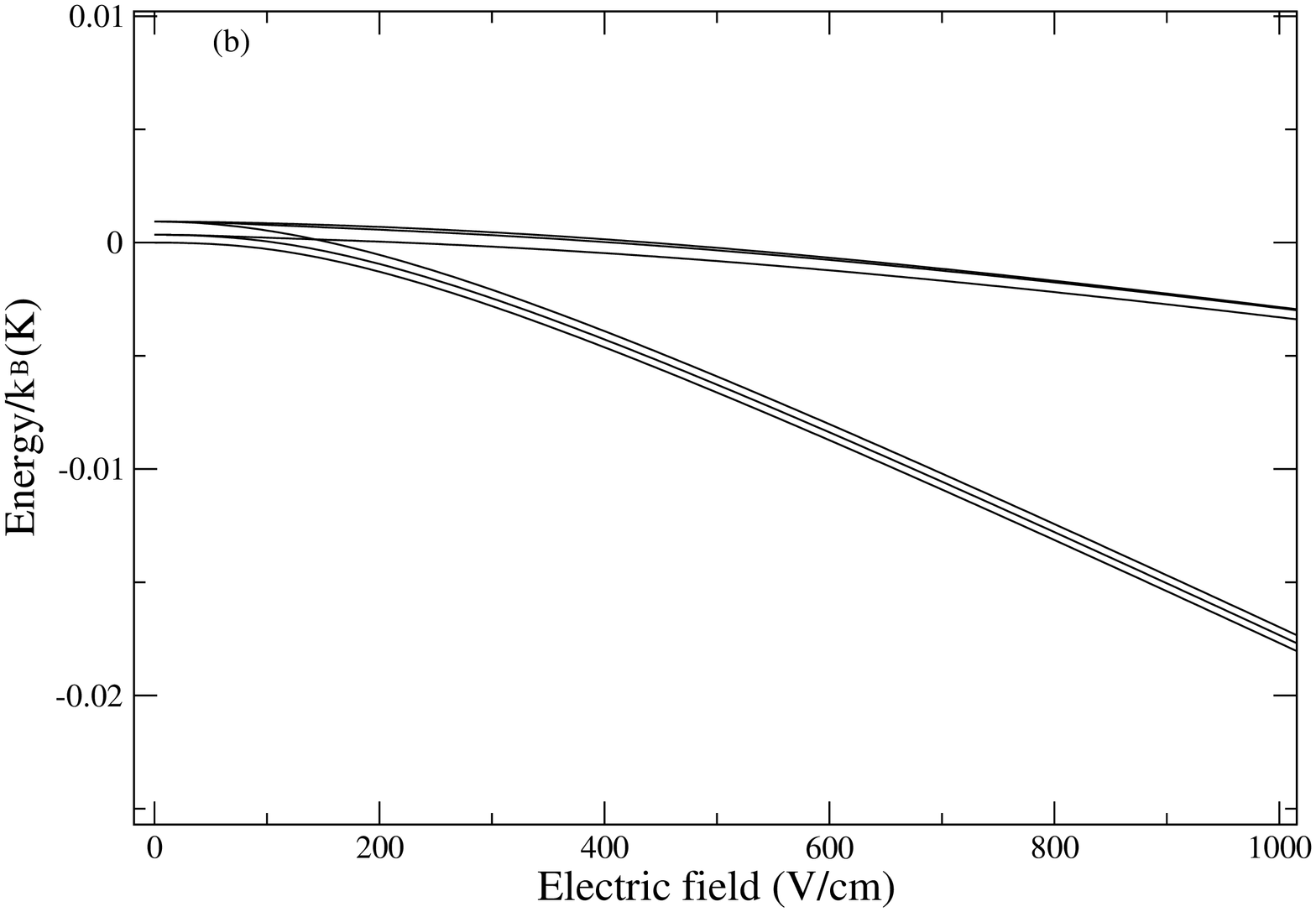}}
\caption{Stark effect for the $^2\Pi_{3/2}|f,m_f,{\rm parity} \rangle$
ground states of OD.  Shown are state of f parity (a) and e parity (b).}
\end{figure}

We consider collisions of the highest-energy weak-field seeking state in 
Fig.1, with quantum numbers $|f,m_f, {\rm parity} \rangle =$
$|5/2,5/2,f \rangle$.
The details of our scattering theory have been presented elsewhere, 
for OH \cite{Avdeenkov1}.  The
main difference in handling OD is to incorporate Fermi exchange symmetry,
which amounts to changing plus signs to minus signs in Eqn. (17)
of Ref. \cite{Avdeenkov1}.
Otherwise, we treat the scattering in the same
way, by including only the Stark and dipole-dipole interactions, along
with the hyperfine structure.

\begin{figure}
\centerline{\includegraphics[width=0.9\linewidth,height=1.08\linewidth,angle=-90]{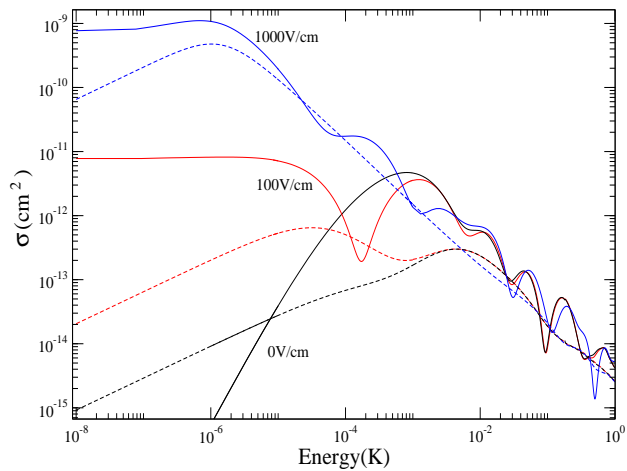}}
\caption{Elastic (solid lines) and total inelastic(dashed) cross sections
for several values of electric field: ${\cal E}=0,100,1000 V/cm$.}
\end{figure}

Fig. 2 shows the main scattering results, as collision cross sections 
versus collision energy at different electric fields.  Three different
applied electric fields are indicated by color coding: ${\cal E}=0$ (black),
 ${\cal E}=100$ V/cm (red), and ${\cal E}=1000$ V/cm (blue).
In each case, the solid line denotes elastic scattering, while
the dashed line represents total inelastic scattering to channels
where one or both molecules loses internal energy.
In zero field, the molecules are completely
unpolarized, and the dipole-dipole interaction vanishes.  Thus the 
cross sections obey the familiar Wigner threshold laws for short-ranged 
interactions between fermions:
the elastic cross section $\sigma_{el} \propto E^2$, whereas the 
(exothermic) inelastic cross section $\sigma_{inel} \propto E^{1/2}$
\cite{Sadeghpour}.  Thus in zero field elastic scattering is 
actually {\it less} likely than inelastic scattering at lower 
energies (below about 10$\mu$K in this example).  Above this energy
elastic scattering appears somewhat more favorable than inelastic 
scattering, at least until
several mK, where both cross sections start to hit the unitarity limit.  

Turning on the electric field partially polarizes the molecules, so that the 
dipole-dipole interaction is ``activated.''  Then the dipole-dipole
threshold laws take effect:
$\sigma_{el} \propto const.$, whereas we still have 
$\sigma_{inel} \propto E^{1/2}$.  Fig.2
illustrates where this threshold behavior kicks in for different electric field
values.  Notice that the higher the electric field, the lower is the 
energy where the threshold behavior is attained.
This is because the effective $C_3$ coefficient that determines 
the barrier height $E_b$ is an increasing function of
electric field, at least until it saturates \cite{Avdeenkov1}.

\begin{figure}
\centerline{\includegraphics[width=0.9\linewidth,height=1.08\linewidth,angle=-90]{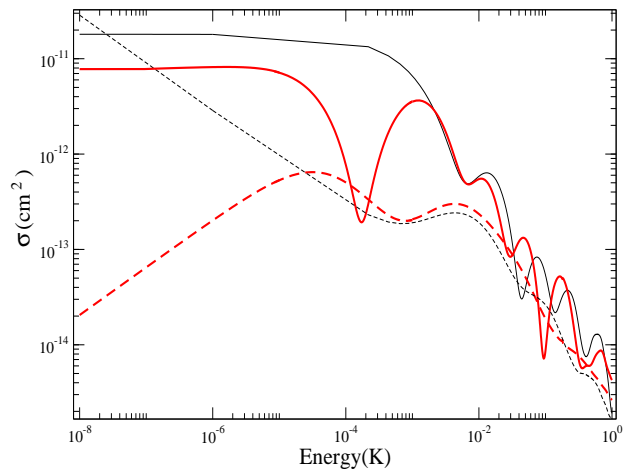}}
\caption{Comparison of cross sections for OD (red) and OH 
(black) molecules.
Solid and dashed lines refer to elastic and inelastic cross sections,
respectively.  }
\end{figure}

On the other hand, a Fermi gas of molecules that is {\it already
cold} will enjoy the benefits of Wigner-law suppression of
inelastic collisions.  Suppose a quantum degenerate gas of
OD could be produced at nK temperatures, as is the case
for current experiments in $^{40}$K and $^6$Li.  Then Fig. 2
suggests that a small bias field of $\sim 100$ V/cm reduces
inelastic cross sections to an acceptable level of 
$\sim 2 \times 10^{-14}$ cm$^2$, 
corresponding to a rate constant $\sim 10^{-16}$ cm$^3$/sec.

To emphasize the difference between bosons and fermions, 
we reproduce the ${\cal E}=100$ V/cm cross sections 
in Fig. 3 (red) along with the corresponding
cross sections for OH in the same field (black).  It is clear that in
both cases elastic scattering (solid lines) is quite similar,
whereas the behavior of inelastic scattering is utterly
different at low energies for the two species.  Equally
importantly, at collision energies above about 1 mK, all the cross
sections have the same general behavior.  This is a manifestation
of the strength of the dipolar interactions, and the fact that
in this energy range all processes are essentially unitarity-limited.

\section{On the question of field-linked resonances}

Finally, we comment on the occurrence of field-linked (FL) resonance 
states in this system.  Fig. 4a shows the elastic and inelastic 
cross sections versus electric field, at a fixed collision energy of 
1 $\mu$K.  This figure exhibits the characteristic peaks indicative
of field-linked resonances; compare Fig.(2) of Ref. \cite{Avdeenkov1}.  
To converge these results at higher field demands an increasing number
of partial waves.  Fig. 4b illustrates the convergence of the 
resonant scattering cross section for various numbers of partial
waves included.  For partial waves $L = 1, 3, 5, 7$, the cross
section is well-converged up to several hundred V/cm.  This is
sufficient to compute the first two resonance states, which are the only
well-resolved ones anyway. 

\begin{figure}
\centerline{\includegraphics[width=0.9\linewidth,height=1.08\linewidth,angle=-90]{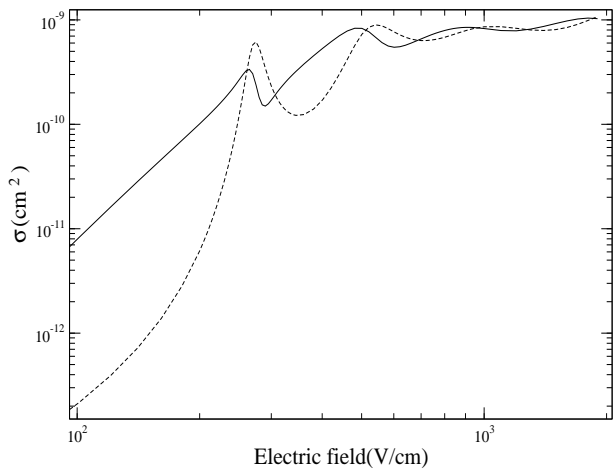}}
\centerline{\includegraphics[width=0.9\linewidth,height=1.08\linewidth,angle=-90]{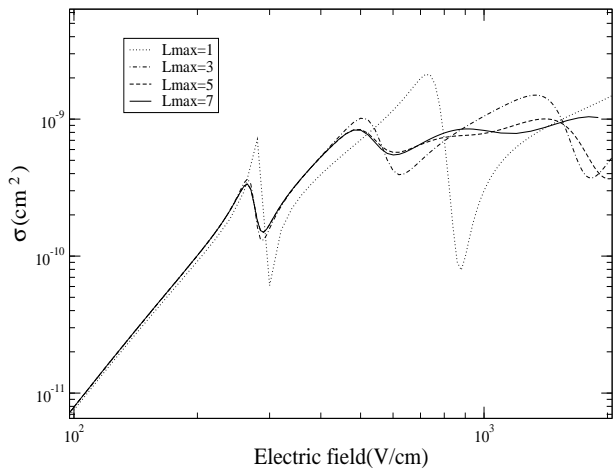}}
\caption{a) Elastic (solid) and inelastic (dashed) cross sections
for OD scattering as a function of applied electric field.  The 
Collision energy is $E=1\mu$K.  b) Convergence of elastic cross
section upon increasing the number of partial waves included in the
calculation.}
\end{figure}

As discussed in Ref. \cite{Avdeenkov1} and elaborated on in 
Ref. \cite{Avdeenkov4}, the FL resonances live
in adiabatic potential energy surfaces generated by avoided crossings in the
long-range dipole-dipole interaction.  The surfaces have  a somewhat different
character for fermions than for bosons, however.  Fig. 5 shows a set 
of adiabatic curves for a single partial wave $L=1$; this is the 
analogue of Fig. (1) in Ref. \cite{Avdeenkov3}. 

\begin{figure}
\centerline{\includegraphics[width=0.9\linewidth,height=1.08\linewidth,angle=-90]{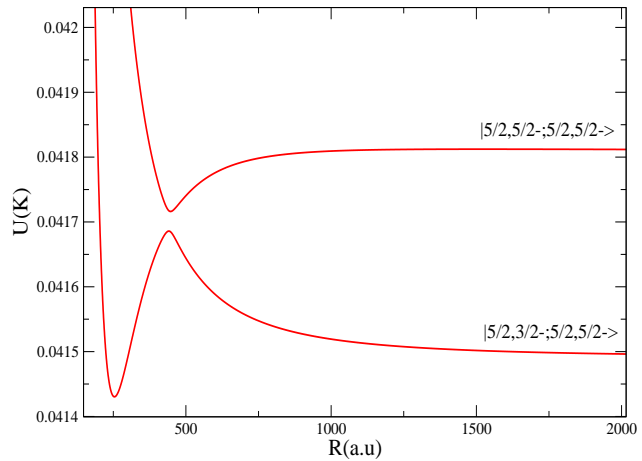}}
\caption{A set of avoided crossings that generate FL states for OD.}
\end{figure}

For identical bosons, the long-range attractive part of the relevant curve is 
predominantly s-wave in character, hence has a $1/R^4$ behavior in an 
electric field.  This reflects the fact that the direct dipolar 
interaction vanishes for s-waves, and makes an effect only at 
second order \cite{Avdeenkov1}.  For fermions in identical spin states,
however, the attractive part involves the p-wave interaction, where 
the dipole is already nonzero, so that the long-range interaction 
scales as $1/R^3$.   The net effect is that 
the inner turning point of the 
s-wave FL states approaches smaller $R$ as the field is increased for bosons,
but that this inner turning point is relatively fixed for fermions.  (The outer
turning point is set by the energy of the resonant state relative to the 
threshold, and is thus arbitrarily large.)  

These resonant states, if sufficiently stable, may form a novel kind of pair of
fermions, which may ultimately lead to an exotic Fermi superfluid state.  
Unfortunately, as seen in Fig. 4, these resonances are quite readily susceptible
to predissociation, indicated by the large inelastic cross sections near resonance.
In this they resemble their bosonic counterparts.  However, stabilization of
cold dipolar gases using magnetic fields has been recently discussed 
\cite{Ticknor}.
It is yet conceivable that these resonances could be tamed long enough to
put them to use.  

In summary, we have computed scattering cross sections for cold 
collisions of the fermionic free radical OD, as functions of both 
collision energy and electric field.  We find that, similar to the 
case of bosonic OH, these molecules are unlikely to be stable against 
collisions in traps warmer than about 10 $\mu$K.  Unlike OH, however,
they will be collisionally  stable at lower temperatures, owing to 
the unique Wigner threshold laws for fermionic polar particles.  
in such a gas.

This work has been supported by the NSF and by a grant from the 
W. M. Keck Foundation.  We acknowledge useful interactions with
M. Kajita and D. DeMille.

\end{document}